\documentstyle[epsfig,aps,floats]{revtex}

\begin{document}
\twocolumn[\columnwidth\textwidth\csname@twocolumnfalse\endcsname

\title{The Density Matrix Renormalization Group Method for
Realistic Large-Scale Nuclear Shell-Model Calculations}

\author{S.S. Dimitrova$^{1,2}$, S.~Pittel$^2$, J. Dukelsky$^3$, and M.V. Stoitsov$^{1,4}$ \\ ~}

\address{$^1$Institute for Nuclear Research and Nuclear Energy, Bulgarian
Academy of Sciences, \\ Sofia-1784, Bulgaria, \\
 $^2$Bartol Research Institute, University of Delaware, \\ Newark, Delaware 19716, USA\\
 $^3$ Instituto de Estructura de la Materia, Consejo
Superior de Investigaciones Cientificas, Serrano 123, \\ 28006
Madrid, Spain}
\address{$^4$Joint Institute for Heavy Ion
Research, Oak Ridge, Tennessee 37831.  Department of Physics,
University of Tennessee, Knoxville, Tennessee 37996.  Physics
Division, Oak Ridge National Laboratory, Oak Ridge, Tennessee
37831}

\maketitle \vspace{7mm}
\begin{abstract} The Density Matrix Renormalization Group
(DMRG) method is developed for application to realistic nuclear
systems. Test results are reported for $^{24}Mg$.

\end{abstract}


\addvspace{12mm}] \narrowtext

\section{Introduction}
The nuclear shell model \cite{R1} is one of the most extensively
used  methods for a microscopic description of the nuclear
structure. Within this approach, the nucleus is treated as an
inert doubly--magic core and a number of valence nucleons,
scattered by effective interaction over an active valence space
consisting of at most a few major shells. Despite the enormous
truncation inherent in this approach, the shell-model method as
just described can still only be applied in very limited nuclear
regimes, namely for those nuclei with a sufficiently small number
of active nucleons or a relatively low degeneracy of the valence
shells that are retained. The largest calculations that have been
reported to date are for the binding energies of nuclei in the
$fp$--shell through $^{64}Zn$~\cite{R2}.

For heavier nuclei or nuclei farther from closed shells, one is forced to make
further truncations in order to reduce the number of shell-model configurations
to a manageable size. The most promising approach now in use is to truncate on
the basis of Monte Carlo sampling\cite{R3}. In this way, it has recently proven
possible to extend the shell model beyond the $fp$--shell to describe the
transition from spherical to deformed nuclei in the Barium isotopes~\cite{R4}.

Nowadays, the Density Matrix Renormalization Group (DMRG) is
recognized as a potentially promising tool for application to
large-- scale nuclear structure calculations. The method was
initially developed and applied in the framework of
low--dimensional quantum lattice systems \cite{R5} and then
subsequently extended to finite Fermi systems to treat a pairing
problem of relevance to ultrasmall superconducting grains
\cite{R6}. This new approach, referred as the particle--hole
(p--h) DMRG, was recently applied to a first test problem of some
relevance to nuclear structure \cite{R7,R8,R9}. The application
involved identical nucleons moving in a large single $j$--shell
under the influence of a pairing plus quadrupole interaction with
an additional single-particle energy term that split the shell
into degenerate doublets. Comparing with the results of exact
diagonalization, it was shown that the method leads to extremely
accurate results for the ground state and for low--lying excited
states without ever requiring the diagonalization of very large
matrices. Furthermore, even when the problem was not amenable to
exact solution, the method was seen to exhibit rapid exponential
convergence. All of this has encouraged us to begin considering
the application of the DMRG method in realistic shell-model
calculations. We report here the results of our first attempt, a
calculation for the nucleus $^{24}Mg$. Since exact shell model
results exist for this nucleus, these calculations provide a
meaningful test of the ability of the p-h DMRG method to work in
realistic nuclear scenarios.

The paper is organized as follows. In Section II, we review the
basic features of the p-h DMRG method. In Section III, we report
results of calculations for a system of 40 like fermions in the
$j=99/2$ shell, the starting point of our recent activities, and
then present the first realistic application of the method to
$^{24}Mg$. Finally in Section IV we summarize our principal
conclusions and outline future directions of the project.

\section{The DMRG procedure}

The basic idea of the DMRG method is to {\it systematically} take into account
the physics of {\it all} single--particle levels. This is done by first taking
into account the most important levels, namely those that are nearest to the
Fermi surface, and then gradually including the others in subsequent
iterations. At each step of the procedure, a truncation is implemented both in
the space of particle states and in the space of hole states, so as to
optimally take into account the effect of the most important states for each of
these two subspaces of the problem. The calculation is carried out as a
function of the number of particle and hole states that are maintained after
each iteration, with the assumption that these numbers are the same. This
parameter, which we will call $p$, is gradually increased and the results are
plotted against it. Prior experience from other applications of the methodology
suggests that the results converge exponentially with $p$. Thus, when we
achieve changes with increasing $p$ that are acceptably small we simply
terminate the calculation.

Since the p--h DMRG procedure has been discussed in some detail
and generality in \cite{R9}, here we just sketch the key steps and
spell out how they are implemented specifically for $^{24}Mg$.
\vspace{0.1cm}

 $1.$~{\it We start by choosing the basis of the problem
and the Fermi level for the nuclear system under consideration.}
$^{24}Mg$ can be considered as a double--magic $^{16}O$ core plus
four valence neutrons and four valence protons, scattered over the
orbits of the $sd$--shell. These are the $1d_{5/2}$, $2s_{1/2}$
and $1d_{3/2}$ levels, with degeneracies $6$, $2$ and $4$,
respectively.

\vspace{0.1cm}

$2.$~{\it The next step is to define the Hamiltonian of the system
in the restricted set of active single-particle states.} The
Hamiltonian contains one-- and two--body terms for like particles
parts $H^{\tau}$, ${\tau}={\nu,\pi}$ and a two--body term for the
proton--neutron part $H^{\nu \pi}$:
\begin{equation}
    H = H^{\nu} + H^{\pi} + H^{\nu \pi}\label{ham1}
\end{equation}

where

\begin{eqnarray}
H^{\tau} &=& \sum_{\alpha m} \epsilon_{\alpha m} a^{\tau \;\dagger}_{\alpha m} a^{\tau}_{\alpha m}  \nonumber \\
&+& \frac{1}{4} \mathop{\sum_{\alpha_1m_1,
\alpha_2m_2}}_{\alpha_3m_3, \alpha_4m_4}
~ \langle \alpha_1m_1,~\alpha_2m_2|V| \alpha_3m_3,~\alpha_4m_4 \rangle  \nonumber \\
&& ~~~~~~~~~~~~~~~~~ \times a^{\tau \; \dagger}_{\alpha_1m_1}
a^{\tau \; \dagger}_{\alpha_2m_2} a^{\tau}_{\alpha_4m_4}
a^{\tau}_{\alpha_3m_3} \label{ham2}
\end{eqnarray}

and

\begin{eqnarray}
H^{\nu \pi} &=& \mathop{\sum_{\alpha_1m_1,
\alpha_2m_2}}_{\alpha_3m_3, \alpha_4m_4}
~ \langle \alpha_1m_1,~\alpha_2m_2|V| \alpha_3m_3,~\alpha_4m_4 \rangle  \nonumber \\
&& ~~~~~~~~~~~~~~~~~ \times a^{\nu\;\dagger}_{\alpha_1m_1}
a^{\pi\;\dagger}_{\alpha_2m_2} a^{\nu}_{\alpha_4m_4}
a^{\pi}_{\alpha_3m_3} \; . \label{pnham}
\end{eqnarray}

\vspace{0.05cm} Steps 1 and 2 together define the shell--model
problem.

\vspace{0.1cm} $3.$~{\it The next step is to split up the set of
multiply-degenerate spherical shell model levels into an
appropriate ordered set of doubly-degenerate levels, which will be
taken into account iteratively in the p--h DMRG procedure.} In the
case of $^{24}Mg$, the low--lying states are expected to be
prolate deformed. This suggests that we first carry out a Hartree
Fock calculation of $^{24}Mg$, using the chosen shell--model
Hamiltonian,  to define an appropriate prolate--deformed
single--particle basis. The procedure, which is schematically
illustrated in figure 1, leads to a set of doubly-degenerate
levels, each having a definite value of the projection of angular
momentum on the symmetry axis. For $^{24}Mg$, the Fermi energy
both for neutrons and protons is between the first $3/2^+$ level
and the second $1/2^+$ level.

\begin{figure}[htb]
\vspace{-3.5cm}

\begin{center}
\leavevmode \epsfxsize=13.5cm \epsffile{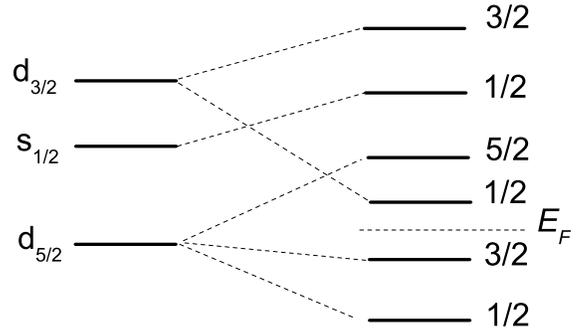}
\end{center}
\vspace{-1.50cm} \narrowtext\caption{\it Schematic illustration of the
splitting of the model--space single--particle levels within the $sd$--shell
into a set of doubly--degenerate levels by an axially--deformed Hartree--Fock
calculation. The dashed line represents the Fermi energy ($E_F$), which
separates the particle levels from the hole levels. Each doubly--degenerate
level is labelled by its angular momentum projection on the intrinsic z-axis.}
\end{figure}

The Fermi surface splits the shell into two kind of states -- the
hole states below the Fermi level and the particle states above
it. According to the p--h DMRG prescription we take first into
account the particle and hole states closest to the Fermi surface
and then gradually involve all of the others that are further
away.

Note of course that for the nucleus $^{24}Mg$  there are four type
of levels - particle and hole levels for neutrons and particle and
hole levels for protons.

\vspace{0.1cm}

$4.$~{\it We initialize the DMRG procedure by considering as
active the lowest particle state above the Fermi surface and the
highest hole state below.} In the case of $^{24}Mg$, this means
taking into account the $3/2^+_1$ hole level and the $5/2^+_1$
particle level, as they are the ones closest to the Fermi surface.
For this set of particle states and hole states for protons and
nucleons, we calculate the hamiltonian matrix and the matrices of
all of its sub-operators, namely $a, \; a a, \; a^{\dag} a, \;
a^{\dag} a^{\dag} a$, and $a^{\dag} a^{\dag} a a$. Thus in a
system of neutrons and protons we have four distinct blocks --
neutron particle, proton particle, neutron hole and proton hole
states.

\vspace{0.1cm}

$5.$~ {\it We then proceed to the first iteration by adding the
next higher particle level and the next lower hole level.} For
$^{24}Mg$, these are the $1/2^+_1$ hole level and the $1/2^+_2$
particle level. In our calculations, we in fact add four levels,
one for proton particles, one for proton holes, one for neutron
particles and one for neutron holes.

We can express the particle and hole states in these enlarged spaces as

\begin{equation}
|\,I\, \rangle ~=~ |\,i\,\rangle_{old}\; |\,j\,\rangle_{new} ~,
\end{equation}
where $|\,i\,\rangle_{old}$ refers to the $4$ particle (hole)
states within the first iteration and $|\,j\,\rangle_{new}$ -- to
the $4$ new states from the additional level. Thus, each of these
four blocks now contains 16 states.

To determine the matrix elements of the hamiltonian and all of its
sub-operators in the proton--particle, proton--hole,
neutron--particle and neutron--hole subspaces, we make use of the
fact that all matrix elements in the {\em old} space are already
known from the previous iteration while those coming from the {\em
new} level are very simple to calculate. For example, the
expectation values of the operators $a^{\dag} a$ in the enlarged
space looks like

\begin{eqnarray}
\langle i, j | a^{\dagger}_{\alpha_1m_1} a_{\alpha_2m_2} | k,l \rangle &&
=\langle i | a^{\dagger}_{\alpha_1m_1} a_{\alpha_2m_2}
| k \rangle ~\delta_{j,l}\nonumber \\
 &&+
\langle j | a^{\dagger}_{\alpha_1m_1} a_{\alpha_2m_2} | l \rangle ~\delta_{i,k} \nonumber \\
&&+ (-)^{n_k} \langle i | a^{\dagger}_{\alpha_1m_1} | k \rangle ~
\langle j | a_{\alpha_2m_2} | l \rangle \nonumber \\
&&- (-)^{n_k} \langle i | a_{\alpha_2m_2} | k \rangle ~ \langle
j | a^{\dagger}_{\alpha_1m_1} | l \rangle ~, \nonumber \\
\label{enlarge}
\end{eqnarray}
where $n_i$ is the number of particles in state $|i \rangle$.

\vspace{0.1cm}

$6.$~ {\it The next step is to couple the states in the four
blocks.} In doing this, we only keep those product states in which
the total number of particles for protons equals the total number
of holes for protons and the same for neutrons (to make the theory
particle number conserving) and in which the total angular
momentum projection of the system is $M=0$. We will call the
number of such coupled states $N$. Note that it is significantly
less than $16 \times 16$ because of the above restrictions on the
number of particles and holes and on the total $M$ value. We then
calculate the matrix elements of the full hamiltonian
eqs.(\ref{ham1},\ref{ham2},\ref{pnham}) in this product basis
(often called the {\em superblock}), making use of the fact that
we know the matrix elements of the hamiltonian and all its
sub-operators separately in the particle and hole spaces for
protons and neutrons.

\vspace{0.1cm}

$7.$~ {\it Next we diagonalize the superblock hamiltonian:}

\begin{equation}
H|\Psi_\alpha \rangle = E_\alpha |\Psi_\alpha \rangle ~,
\end{equation}
with
\begin{equation}
|\Psi_\alpha \rangle =\mathop{
\sum_{i^{\nu}_p,j^{\nu}_h}}_{{k^{\pi}_p,l^{\pi}_h}} \Psi^{(\alpha)}_{ijkl}
|i^{\nu}_p \rangle |j^{\nu}_h \rangle |k^{\pi}_p \rangle |l^{ \pi}_h \rangle ~.
\label{superblock}
\end{equation}
Note that the sums go over the all $16$ states in the respective enlarged
particle and hole blocks for protons and neutrons.

\vspace{0.1cm}

$8.$~{\it The next step is to truncate to the optimum $p$ states
in the four blocks, optimum in the sense that the states we retain
provide the optimum approximation to the system prior to
truncation.}

If $p$ is greater than $16$, no truncation is required and we
simply continue to the next iteration, adding the next levels for
particles and holes.

If $p$ is less than $16$, we perform the optimized truncation in
the following way. If we want to optimize the description of the L
lowest eigenstates of the hamiltonian, we have to construct the
mixed density matrices to these L eigenstates in the four blocks -
particles and holes ($p,h$) for protons and
neutrons($\tau=\nu,\pi$). For example for neutrons they are:

\begin{eqnarray}
\rho^{\nu\;p}_{ii^\prime} = \frac{1}{L} \sum_{\alpha=1,L} \sum_{j,k,l=1,4p}
\Psi^{\alpha }_{ijkl} {\Psi^{\alpha *}_{i^{\prime}jkl}} ~,
\nonumber \\
\rho^{\nu \;h}_{jj^\prime} = \frac{1}{L} \sum_{\alpha=1,L} \sum_{i,k,l=1,4p}
\Psi^{\alpha}_{ijkl} {\Psi^{\alpha *}_{ij^{\prime} kl}}~. \label{rho}
\end{eqnarray}

We then diagonalize all four of these density matrices, each of
which is of dimension $4p$:

\begin{eqnarray}
\rho^{\tau p} \left| \,u^{\beta}\,\right\rangle_{\tau p} ~=~ \omega^{\tau
p}_{\beta} ~ \left| \,u^{\beta}\,\right\rangle_{\tau p} ~,
\nonumber \\
\rho^{\tau h} \left| \,u^{\beta}\,\right\rangle_{\tau h} ~=~ \omega^{\tau
h}_{\beta} ~ \left| \,u^{\beta}\,\right\rangle_{\tau h} ~. \label{rhodiag}
\end{eqnarray}
Those $p$ eigenstates with the largest eigenvalues  provide the
optimum approximation to the set of $L$ states that were targeted
in constructing the corresponding mixed density matrix.

\vspace{0.1cm}

$9.$~{\it The final step of the iteration is to transform the
matrices of all needed combinations of creation and annihilation
operators in the four blocks from the $4p$--dimensional spaces to
the optimal $p$--dimensional truncated spaces.}

\vspace{0.1cm}

$10..$~{\it The next step is to proceed to the next iteration by
adding the next set of levels and following steps $5-9$.} We
continue to add one level from each block, until one (or more) of
them is exhausted. From that point on, we only add levels from the
remaining blocks, and only carry out the optimized truncation for
them. The procedure ends when all states of the four block are
exhausted. Note that in the case of $^{24}Mg$, there are a total
of three iterations. In the first iteration, both particle and
hole levels are added. In subsequent iterations only particle
levels are added.

\section{Results}

Before presenting the results of our realistic calculations for $^{24}Mg$ we
first return for a moment to the single-j shell--model system discussed in
\cite{R8,R9}. The largest calculations we have so far carried out is for a
system of 40 particles occupying a single $j=99/2$ orbit and interacting via
pairing plus quadrupole interaction with an additional one-body term to split
the degeneracy of the orbit.

\begin{figure}[htb]
\vspace{-0.5cm}
\begin{center}
\leavevmode \epsfig{file=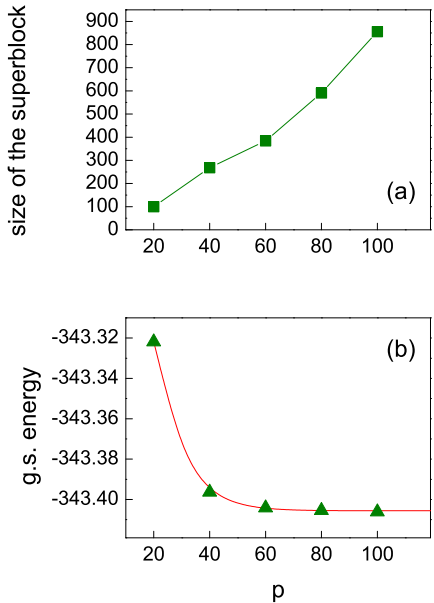,width=6.5cm}
\end{center}

\vspace{-0.70cm} \narrowtext \caption{\it The maximum size of the superblock
(a) and the ground-state energy (b) for a system of 40 identical nucleons in a
single $j=99/2$ orbit interacting via pairing plus quadrupole Hamiltonian. An
exponential fit to these results is also plotted. }
\end{figure}

In this case, the exact calculation would involve a hamiltonian matrix of
dimension $3.84007 \times 10^{25}$, obviously much too large to treat without
dramatic truncation. In figure 2, we display the largest size of the
hamiltonian matrix we had to diagonalize and the ground state energy of the
system as a function of $p$. It is seen that while the energy of the ground
state follows a steep exponential trend, the size of the superblock increases
linearly. This gives us hope that we can treat realistic nuclear systems
accurately using the DMRG strategy, while keeping the size of the matrices
manageable.
\begin{figure}[htb]
\vspace{-1.0cm}
\begin{center}
\leavevmode \epsfig{file=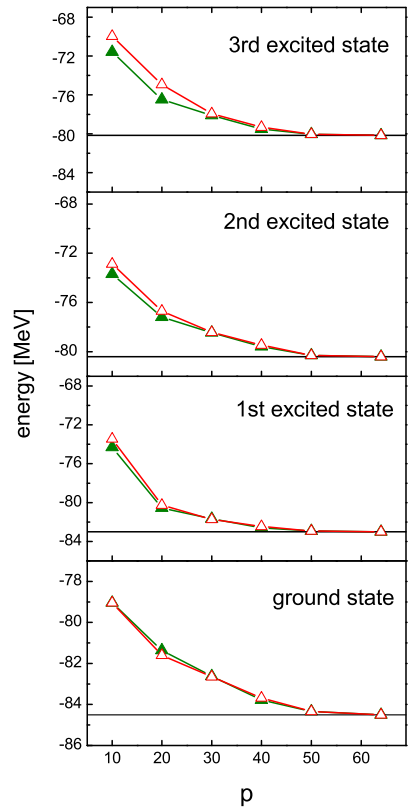,width=6.5cm}
\end{center}

\vspace{-0.70cm} \narrowtext \caption{\it The energy of the ground state and of
the three lowest excited states for $^{24}Mg$. The open triangles are the
results when only the ground state is targeted in the optimization procedure;
the solid triangles refer to calculations when the lowest four states of the
system are all targeted simultaneously.}
\end{figure}

\begin{figure}[htb]
\vspace{-1.0cm}
\begin{center}
\leavevmode \epsfig{file=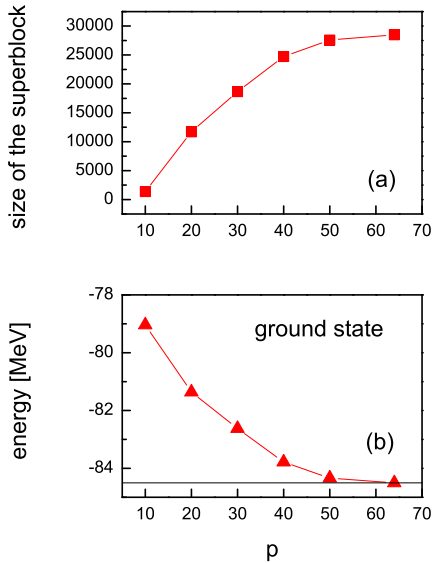,width=6.5cm}
\end{center}
 \vspace{-0.70cm} \narrowtext \caption{\it The maximum size of the superblock $(a)$ and the energy of the ground
 state $(b)$ for $^{24}Mg$, when only the ground state is targeted in the optimization
procedure. }
\end{figure}

\begin{figure}[htb]
\vspace{-1.50cm}
\begin{center}
\leavevmode \epsfig{file=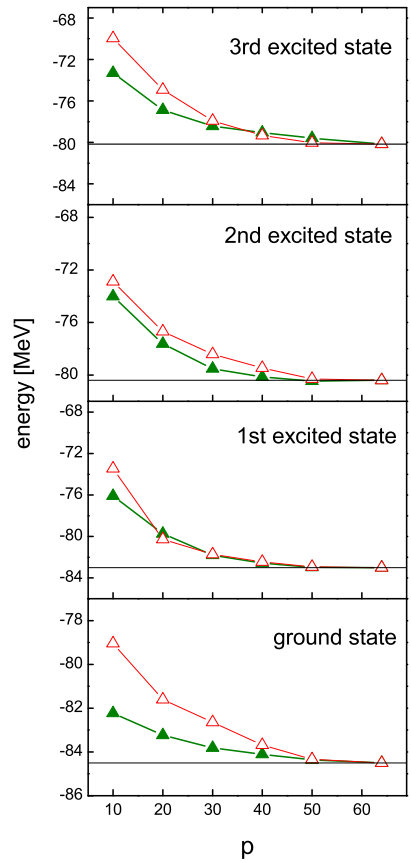,width=6.5cm}
\end{center}

\vspace{-0.70cm} \narrowtext \caption{\it The energy of the ground state and of
the three lowest excited states for $^{24}Mg$, when only the ground state is
targeted in the optimization procedure. The solid triangles refer to
calculations in the deformed HF-basis and the open triangles to calculations in
the spherical basis.}
\end{figure}

The first realistic p-h DMRG calculations we performed were for $^{24}Mg$. As
noted earlier, this system is assumed in the shell model to be a $^{16}O$ core
plus four valence neutrons and four valence protons in the $sd$--shell. We have
used the Wildenthal's {\it USD} interaction \cite{R10,R11}. The main reason for
considering this nucleus first is that the shell--model problem for $^{24}Mg$
can be solved exactly. The size of the Hilbert space in the $m$--scheme is
$28,503$ for which the Hamiltonian matrix can be treated using the Lanczos
algorithm.

In these calculations, we have considered two possible strategies for defining
the order of doubly--degenerate levels to include in the iterative DMRG
procedure. As discussed in Section II, the most appropriate strategy is to use
an axially-symmetric HF calculation to define the states and to calculate all
matrix elements in that deformed basis. A simpler strategy is to use the HF
procedure to tell us the order in which to fill the $|m|$ values, but to still
carry out the calculation in the spherical basis. For $^{24}Mg$, this would
correspond to an ordering of levels such that the $d_{5/2~1/2}$ is lowest,
followed by the $d_{5/2~3/2}$, the $d_{3/2~1/2}$, the $d_{5/2~5/2}$, the
$s_{1/2~1/2}$ and then finally the $d_{3/2~3/2}$. The latter strategy should
converge more slowly, but is simpler to implement and less time consuming.

We first show the results using a spherical shell--model basis,
but with a slightly different ordering of levels than above,
namely $d_{5/2~1/2}$, $d_{5/2~3/2}$, $d_{5/2~5/2}$, $s_{1/2~1/2}$,
$d_{3/2~1/2}$ and $d_{3/2~3/2}$ . Figure 3 shows the ground state
energy and the energy of the first three excited states as a
function of the number of states $p$ kept for each block during
the DMRG procedure. The exact results are represented by a
straight line. For $p=64$ the whole shell--model space is
exhausted and the exact results are reproduced. Results for two
set of calculations are displayed -- when just the ground state is
taken into account in the reduced density matrices
(eqs.[\ref{rho}]) and when the lowest four states are targeted
simultaneously. A first look at this figure tells us that,
contrary to our expectations, the exponential conversion of the
energy with $p$ is not especially rapid, neither for the ground
state nor for the excited states. From figure 4, we see that an
accuracy of $0.2\%$ for the ground state energy requires $p=50$,
where $97\%$ of the shell-model states are taken into account.
Moreover the size of the superblock increases exponentially with
$p$, contrary to the single--j case where the dependence was
linear.

Going back to figure 3, we also see that including excited states
in defining the reduced density matrices improves slightly the
description of the energy of the excited states without
significantly changing the ground state energy.  Moreover once $p$
becomes large enough, there is no discernable difference between
the two sets of results.

Next we consider what happens when the calculations are carried out in the
deformed Hartree--Fock single--particle basis. Figure 5 compares the results
obtained for the ground state energy and the energies of the three lowest
excited states as a function of $p$ in the spherical basis (the open triangles)
and in the HF--basis (the full triangles). Is is seen that there is a
significant improvement in the results, especially for the ground state energy,
for small values of $p$. For $p$ values larger then $40$, however, use of the
HF--basis is of no great value. Most importantly, in neither case can we
achieve a high level of accuracy without including a large fraction of the full
Hilbert space.

\section{Closing Remarks}

In this paper, we presented results of the first p-h DMRG calculations carried
out for a realistic nuclear system. We considered the nucleus $^{24}Mg$, for
which exact shell--model calculations assuming an inert $^{16}O$ core and 8
valence nucleons in the $sd$--shell have been reported. Our calculations used
the same Wildenthal $USD$--interaction as the exact calculations, so as to
permit a meaningful test of the DMRG method.

The results show that, independent of the single--particle basis
used, the exponential convergence for the ground state energy and
for the energies of the lowest excited states is fairly slow. To
get accurate results we must include almost the complete space.

The first question to be addressed in the future is whether these
results are a consequence of the very small shell--model space for
$^{24}Mg$. We will thus consider the somewhat larger, but still
exactly solvable, problem of $^{48}Cr$.

The next step after that is to include {\it sweeps} in the DMRG
method, whereby we go through the set of levels several times
\cite{R5}. While this will no doubt lead to better accuracy of the
method, it still remains to be seen whether it will lead to very
accurate results with a relatively small number of states kept. If
so, we will then turn to the ultimate goal of this project, to use
the method to treat larger--scale and more challenging nuclear
structure problems.

 \acknowledgments {This work was supported in part by the National
Science Foundation under grant \# PHY-9970749, by the Spanish DGI
under grant BFM2000-1320-C02-02,  by NATO under grant
PST.CLG.977000, by the Bulgarian Science Foundation under contract
$\Phi-905$ and by the Bulgarian--Spanish Exchange Program under
grant \# 2001BG0009. One of the authors (SSD) would also like to
acknowledge the partial support of a Fulbright Visiting Scholar
Grant and the hospitality of the Bartol Research Institute and the
University of Delaware where much of this work was carried out.
Finally, SSD and MVS acknowledge valuable discussions with David
Dean and Thomas Papenbrock.}

\end{document}